\documentclass[twocolumn,secnumarabic,amssymb, nobibnotes, aps, prd, superscriptaddress]{revtex4}
\usepackage{color}

\usepackage[utf8]{inputenc}
\usepackage{graphics,graphicx}
\usepackage{amsmath,amssymb,amsfonts,latexsym,cancel}
\usepackage{bm}

\begin{document}

\title{Fluid pulsation modes from strange stars in a higher-dimensional space-time}
\author{Jos\'e D. V. Arba\~nil}
\email{jose.arbanil@upn.pe}
\affiliation{Departamento de Ciencias, Universidad Privada del Norte, Avenida el Sol 461 San Juan de Lurigancho, 15434 Lima,  Peru}
\affiliation{Facultad de Ciencias F\'isicas, Universidad Nacional Mayor de San Marcos, Avenida Venezuela s/n Cercado de Lima, 15081 Lima,  Peru}
\author{C\'esar H. Lenzi}
\email{chlenzi@ita.br}
\affiliation{Departamento de F\'isica, Instituto Tecnol\'ogico de Aeron\'autica, Centro T\'ecnico Aeroespacial, 12228-900 S\~ao Jos\'e dos Campos, S\~ao Paulo, Brazil}
\author{Manuel Malheiro}
\email{malheiro@ita.br}
\affiliation{Departamento de F\'isica, Instituto Tecnol\'ogico de Aeron\'autica, Centro T\'ecnico Aeroespacial, 12228-900 S\~ao Jos\'e dos Campos, S\~ao Paulo, Brazil}

\date{\today}

\begin{abstract}
In this work, we make the first step to derive non-radial pulsation equations in extra dimensions and investigate how the $f$- and $p_1$-mode frequencies of strange quark stars, within the Cowling approximation, change with the number of dimensions. In this regard, the study is performed by solving numerically the non-radial pulsation equations, adjusted for a $d$-dimensional space-time $(d\geq4)$. We connect the interior to a Schwarzschild-Tangherlini exterior metric and analyze the $f$- and $p_1$- mode frequencies. We found that the frequencies could become higher than those found in four-dimensional space-time. The $f$-mode frequency is essentially constant and only for large gravitational radius values grows monotonically and fast with the gravitational radius. In a gravitational radius range, where $f$-mode frequencies are constant, they increase for space-time dimensions $4\leq d\leq6$ and decrease for $d\geq7$. Regarding $p_1$-mode frequencies they are always larger for higher dimensions and decay monotonically with the increase of the gravitational radius. In extra dimensions, as it happens for four-dimensional space-time, we found $p_1$-mode frequencies are always larger than the $f$-modes ones. In the Newtonian gravity, for a homogeneous star in $d$ dimensions, we observe that the $f$-mode eigenfrequencies are constant and given by the relation $\omega^2=l\, M\, G_d/R^{d-1}$; where $l$ represents the spherical harmonic index, $M\,G_d$ being the total star mass and $R$ the stellar radius. For some gravitational radius interval, we show that a homogeneous star in Newtonian gravity is a good approximation to investigate the $f$-mode frequency of strange stars in the relativistic frame.  In each dimension considered, we find that the $f$-mode frequencies of strange stars are essentially constant since they depend on the average star energy density that is almost constant as a function of the total star mass. Moreover, for a fixed energy density, we also find that the $f$-mode frequency changes with the volume of the unitary sphere in $d-1$ dimension, which attains its maximum value at $d=6$. In neutron stars in four-dimensions, where the average energy density of the star increase with the central energy density, the $f$-mode frequencies will increase with the star mass. Thus, the possibility of measure in gravitational wave detectors the $f$-mode oscillation frequency coming from compact stars with different pulsar masses and observe almost constant frequency values, for $d=4$, in the range $f\sim2-3\,[\rm kHz]$ with $ M\leq1.8M_\odot$, it would be a good sign of the existence of strange quark stars that still lack an astronomical confirmation. Finally, if the $f$-mode frequencies are still constant and different from the range values of $d=4$ for larger total masses, it would be evidence that quarks can propagate in extra space-time dimensions and strange quark stars in $d$ dimension could exist.  
\end{abstract}
\maketitle


\section{Introduction}\label{sec-introd}
\

The observation carried out by the LIGO and Virgo collaborations, the Gravitational Waves (GWs) from a binary black holes merger \cite{abbott2016,abbott2016_2,abbott2017}, turn out to be consistent with foretold from general relativity theory. Shortly after this measurement, GWs coming from a merging neutron star pair was also detected \cite{abbott2017_2}. The latter event, together with its electromagnetic counterpart \cite{abbott2017_3}, has paved a new route on fundamental physics to explore the astrophysical observation. 

Few years after Einstein's theory of gravity was established and influenced by the idea of a generalized theory of gravitation -which, in principle, would unify gravitational and the electromagnetic force- Kaluza \cite{kaluza1921} and Klein \cite{klein1926} argued that extra dimensions could be the route toward a unified theory of gravitation. Many decades after this proposal and motivated by the idea that GWs could assist to prove the existence of extra dimensions, it has been a great deal of interest in the physics community for studying various astrophysical phenomena in extra dimensions. 

Within this gravity theory context, aiming to understand the extra-dimensional space-time effect of some phenomena on compact objects research, several theoretical investigations have been carried out. For instance, the implications of space-time dimension on static equilibrium configurations \cite{harko_mak2000,poncedeleon2000,blazquez2019}, radial stability \cite{arbanil_malheiro2019}, compactness \cite{paul2001,dadhichPRD,chakraborty2017} and gravitational collapse \cite{ghosh_beesham2001,ghosh_beesham2000,ghosh_dadhich2001,debnath2004} of compact stars have been widely investigated. 

In particular, Refs.~\cite {harko_mak2000, poncedeleon2000} investigated the equilibrium configurations in higher dimensional space-time, by solving the stellar structure equations. Thus, considering a hypersphere composed by an incompressible fluid, it has been found that the extra dimensions increase the total mass of compact objects. In this regard, in Ref.~\cite{blazquez2019} a comparative analysis of Boson and Dirac stars in a $d$-dimensional space-time has been investigated. There, the authors found that the solutions of gravitating matter systems depend on the number of dimensions. 

Inspired in the aforementioned works, within the radial perturbation approach, Ref.~\cite{arbanil_malheiro2019} investigated the stability of compact fluid spheres in an extra-dimensional space-time. Considering a hypersphere composed of a perfect fluid that follows the MIT bag model equation of state, it has been shown that extra dimension affects the radial stability of these objects. For a range of central energy densities and total masses, it is shown that the radial stability of strange stars increases with the dimension. Strange quark stars are a possible structure for compact stars but quite different from neutron stars since they are essentially self-bound stars, can have smaller radii, and for bare quark stars, they present a sharp surface where the energy density is not zero. In the MIT quark model, this energy density at the star surface is directly related to the bag constant (the deconfinement energy). These compact objects and their radial stability have been studied \cite{malheiro2/2003,malheiro4/2007}, and also anisotropic pressures and charge effects were investigated \cite{malheiro3/2016,negreiros/2009,malheiro0/2015}. As mentioned in \cite{arbanil_malheiro2019}, the stability of these stars in a  higher dimensional space-time could be associated with the fact that the deconfinement density energy is present. These all stable equilibrium solutions are far from reaching the Buchdahl limit \cite{buchdahl} for a $d$-dimensional space-time \cite{paul2001,dadhichPRD,chakraborty2017}. Thus, it is important to investigate the fluid pulsation modes of these $d$-dimensional strange stars, and see how their frequencies change when the space-time dimensions increase. 

The Buchdahl bound for the $d$-dimensional case states that the radius-mass ratio, i.e., the reason between $R$ and $M\,G_d/(d-3)$, of a compact object follows the inequality $(d-3)R^{d-3}/MG_d\geq (d-1)^{2}/2(d-2)$ \cite{paul2001,dadhichPRD,chakraborty2017}.
 Certainly, a compact object that violates this bound would result in a gravitational collapse. The end outcome of a gravitational collapse depends on the initial conditions imposed, its final stage could be either a naked singularity or a black hole \cite{ghosh_beesham2001,ghosh_beesham2000,ghosh_dadhich2001,debnath2004}. It is worth mentioning that the influence of extra dimension  on some black holes properties have been also investigated; see, e.g., \cite{ishibachi2003,cardoso2003,cardoso2004,myers_perry1986,emparan_reall2008,gibbons_ida_shiromizu,blazquez2018,andre2019}.

Although the influence of the extra dimensions was addressed in the aforementioned contexts, it is of interest to see, within the Cowling approximation, how these could influence the fluid pulsation modes from stable strange stars in higher-dimensional space-time (review \cite{arbanil_malheiro2019}). In our knowledge, it is the first time that the nonradial pulsation equation in $d$ space-time dimensions are obtained, in Cowling approximation, and solved numerically. For this purpose, we numerically integrate the hydrostatic equilibrium equations \cite{tolman,oppievolkoff} and the nonradial pulsation equations \cite{mcdermott1983,lindblom1990}, properly modified to include the extra dimensions. 

We regard this paper as follows: The general relativistic formulation in higher-dimensional space-time is shown in Sec.~\ref{sec-basicequations}; moreover, the steps to follow to obtain both the stellar structure and the nonradial oscillation equations and the equation of state are displayed. The influence of the space-time dimension on the $f$- and $p_1$-modes frequencies ($f$- and $p_1$-mode for short) are presented in Sec.~\ref{results}. Additionally, in this section, there is also presented the oscillation spectrum in the Newtonian limit. We conclude and make some final remarks in  Sec.~\ref{conclusion}. 
The units in which $c=1=G_4$, with $c$ and $G_4$ being respectively the speed of light and the four-dimensional gravitational constant, is adopted through the paper.

\section{General relativistic formulation in high dimensions}\label{sec-basicequations}

\subsection{Field equation}

The nonradial oscillations of neutron stars in higher dimensions within the Cowling approximation are investigated in the framework of general relativity. For a framework in high space-time dimensions, $d\geq4$, the Einstein field equation can be represented through the equality \cite{lemosezanchinbonnor}:
\begin{equation}\label{einstein_tensor}
R_{\mu\nu}-\frac{1}{2}g_{\mu\nu} R= \frac{d-2}{d-3}S_{d-2}G_d T_{\mu\nu}.
\end{equation}
The left side of Eq.~\eqref{einstein_tensor} contains the Ricci tensor $R_{\mu\nu}$, the Ricci scalar $R$ and the metric tensor $g_{\mu\nu}$. In turn, the right hand side bears the universal constant $G_d$, in $d=4$ it represents to the Newton's gravitational constant, and the area of unitary hypersphere $S_{d-2}=2\pi^{(d-1)/2}/\Gamma((d-1)/2)$, with $\Gamma$ being the usual gamma function and the factor $(d-2)\,S_{d-2}/(d-3)$ corresponds to the $8\pi$ term in four dimensions. Moreover, $T_{\mu\nu}$ depicts the perfect fluid energy-momentum tensor given by:
\begin{equation}\label{fluidemt}
T_{\mu\nu}= \left(\rho_{d}+p_{d}\right)U_\mu U_\nu +p_{d}\,g_{\mu\nu},
\end{equation}
where $\rho_{d}$ and $p_{d}$ are respectively the energy density and the fluid pressure and $U_\mu$ is the velocity of the fluid in a $d$-dimensional space-time, where $U_{\mu}U^{\mu}=-1$.

All aforementioned Greek indexes previously defined $\mu,\nu$, etc., run from $0$ to $d-1$, with $0$ representing the time, and the $d-1$ the spacelike coordinates.

\subsection{Static structure equations for a higher dimensional space-time}

To describe the static perfect fluid hypersphere, the higher-dimensional space-time is assumed to be of the form \cite{arbanil_malheiro2019}:
\begin{equation}\label{int_match}
ds^{2}=-e^{2\nu}\,dt^2+e^{2\lambda}\,dr^2+r^2\sum^{d-2}_{i=1}
\left(\prod^{i-1}_{j=1}\sin^{2}\theta_{j}\right)d\theta^{2}_{i},
\end{equation}
where the functions $\nu=\nu(r)$ and $\lambda=\lambda(r)$ depend on the radial coordinate $r$ only. 

To analyze the stellar equilibrium configurations of compact objects, the stellar structure equations in $d$ dimensions must be resolved. These equations could be expressed through the equalities:
\begin{eqnarray}
&&\frac{dm}{dr}=S_{d-2}\rho_{d} r^{d-2}\,,\label{dm/dr}\\
&&\frac{dp_{d}}{dr}=-(p_{d}+\rho_{d})G_{d}\left[\frac{S_{d-2}p_{d}r}{(d-3)}+\frac{m}{r^{d-2}}\right]e^{2\lambda},\label{tov}\\
&&\frac{d\nu}{dr}=-\frac{1}{(p_{d}+\rho_{d})}\frac{dp_{d}}{dr},\label{df/f}
\end{eqnarray}
where the metric function $e^{2\lambda}$ takes the form
\begin{equation}\label{g11}
e^{2\lambda}=\left(1-\frac{2m\,G_{d}}{(d-3)r^{d-3}}\right)^{-1}.
\end{equation}
Function $m$ denotes the gravitational mass within the hypersphere of radius $r$. Eq.~\eqref{tov} is the Tolman-Oppeheimer-Volkoff (TOV) modified from its original form \cite{tolman,oppievolkoff} to introduce the effects of the extra dimensions \cite{arbanil_malheiro2019}.

The higher dimensional static structure equations are integrated along the radial coordinate, from the center to the surface of the object. In the center of the hypersphere, we adopt the conditions:
\begin{equation}
\begin{array}{l}
m(0)=0,\hspace{1.15cm}\lambda(0)=0,\hspace{1.15cm}\nu(0)=\nu_c,\\ 
p_{d}(0)\,G_d=p_{cd}\,G_d\hspace{0.33cm}{\rm and}\hspace{0.33cm} \rho_{d}(0)\,G_d=\rho_{cd}\,G_d.\label{int_cond}
\end{array}
\end{equation}
The surface of the object $r=R$ is reached when:
\begin{equation}\label{pressure_r}
p_{d}(r=R)\,G_d=0.
\end{equation}
The parameters $p_{c}\,G_d$ and $\rho_{c}\,G_d$ stand the central pressure and the central energy density, respectively.

It is important to mention that, at the object surface, the interior line element matches to the Schwarzschild-Tangherlini vacuum exterior solution \cite{lemosezanchin2009,tangherlini1963}:
\begin{equation}\label{metric_functions}
e^{2\nu(R)}=\frac{1}{e^{2\lambda(R)}}=1-\frac{2M\,G_{d}}{(d-3)r^{d-3}},
\end{equation}
where $MG_d/(d-3)$ represents the total mass of the object.

\subsection{Nonradial oscillations in the Cowling approximation}

The Cowling approximation in a four-dimensional space-time \cite{mcdermott1983,lindblom1990} is employed frequently in literature on compact stars oscillations, e.g., Ref.~\cite{lugones2014,sotani2011}. This approximation helps to simplify equations and, by having a relatively small influence on the solutions, the results obtained are qualitatively correct. Indeed, the Cowling approximation shows a discrepancy of less than $20\%$ and $10\%$ to those obtained respectively by a relativistic numerical approach for $f$ and $p_1$-modes \cite{yoshida1997}. This validates its employment to investigate, for instance, the implication of the rotation rate \cite{stavridis2007,boutloukos2007}, crust elasticity \cite{samuelsson2007} and the internal anisotropy \cite{doneva} in the fluid pulsation modes from compact objects. 

In this subsection, the $f$ and $p_1$-modes from strange stars within the Cowling approximation in a $d$-dimensional space-time are investigated. For such purpose, we keep the metric functions fixed, namely, we consider that the metric perturbation is zero $\delta g_{\mu\nu}=0$ \cite{sotani2011}. The fluid pulsation equations can be derived by considering a variation on the Bianchi identity, $\delta(\nabla_{\mu} T^{\mu\nu})=0$. Keeping the metric functions fixed, we get:
\begin{equation}\label{delta_nabla}
\nabla_{\mu}\left(\delta T^{\mu\nu}\right)=0,    
\end{equation}
with:
\begin{equation}
\begin{split}
\delta T^{\mu\nu}=\left(\delta p_d+\delta\rho_d\right)U^{\mu}U^{\nu}+\left(p_d+\rho_d\right)\left(\delta U^{\mu}\right)U^{\nu}\\
+\left(p_d+\rho_d\right)U^{\mu}\left(\delta U^{\nu}\right)+\delta p_d\,g^{\mu\nu}.
\end{split}
\end{equation}
To investigate what does mean Eq.~\eqref{delta_nabla}, it is necessary to project it orthogonal and along to the $d$-velocity $U^{\mu}$. 

To project Eq.~\eqref{delta_nabla} orthogonally to the $d$-velocity $U^{\mu}$, it is multiplied by the orthogonal tensor $P^{\sigma}_{\nu}=\delta^{\sigma}_{\nu}+U^{\sigma}U_{\nu}$. After some algebra, we derive:
\begin{equation}\label{orthogonal_eq}
\begin{split}
\left(\delta p_d+\delta\rho_d\right)U^{\mu}\nabla_\mu U^{\sigma}+\nabla^{\sigma}\left(\delta p_d\right)+U^{\sigma}U^{\mu}\nabla_{\mu}\left(\delta p_d\right)\\
+\left(p_d+\rho_d\right)U^{\mu}\left(\nabla_{\mu}\left(\delta U^{\sigma}\right)-\nabla^{\sigma}\left(\delta U_{\mu}\right)\right)
=0.
\end{split}
\end{equation}
On the other hand, the projection of Eq.~\eqref{delta_nabla} along the $d$-velocity field $U_{\nu}$ yields:
\begin{equation}\label{along_eq}
\begin{split}
\nabla_{\mu}\left[\left(p_d+\rho_d\right)\delta U^{\mu}\right]+\left(p_d+\rho_d\right)U^{\mu}\nabla_{\mu}U_{\nu}\left(\delta U^{\nu}\right)\\+U^{\mu}\nabla_{\mu}\left(\delta\rho_d\right)=0.
\end{split}
\end{equation}

For the higher-dimensional space-time, it is convenient to consider the Lagrangian fluid displacement vector components ${\bm\varsigma}=(\varsigma^{r},\varsigma^{\theta_1},\hdots,\varsigma^{\theta_{d-2}})$ of the form: 
\begin{eqnarray}
&&\varsigma^r=\frac{e^{-\lambda}}{r^{d-2}}{\tilde Q}Y_l^m,\label{fluid_displacement}\\
&&\varsigma^{\theta_i}=-\frac{\tilde Z}{r^{2}}\left(\prod^{i-1}_{j=1}\frac{1}{\sin^2\theta_j}\right)\frac{\partial Y_l^m}{\partial\theta_i},\label{fluid_displacement2}
\end{eqnarray}
where $i$ goes from $1$ to $d-2$. In Eqs.~\eqref{fluid_displacement} and \eqref{fluid_displacement2}, the functions ${\tilde Q}={\tilde Q}(t,r)$ and ${\tilde Z}={\tilde Z}(t,r)$ depend of both the temporal $t$ and radial coordinate $r$, and $Y_l^m=Y_l^m(\theta_1,\hdots,\theta_{d-2})$ represents the spherical harmonics for a higher-dimensional space-time. Then, the perturbations of the velocity of the fluid in a $d$-dimensional space-time, $\delta U^{\mu}=\left(0, \delta U^{r}, \delta U^{\theta_1},\hdots,\delta U^{\theta_{d-2}}\right)$, can be placed as
\begin{eqnarray}
&&\delta U^{r}=\frac{d\varsigma^{r}}{d\tau}=e^{-\nu}\frac{d\varsigma^r}{dt},\label{ur}\\
&&\delta U^{\theta_i}=\frac{d\varsigma^{\theta_i}}{d\tau}=e^{-\nu}\frac{d\varsigma^{\theta_i}}{dt}.\label{uo}
\end{eqnarray}
It is important to say that Eqs.~\eqref{fluid_displacement} and \eqref{fluid_displacement2} are reduced to the form used in \cite{sotani2011} considering $d=4$. 

Taking into account that $U^{\sigma}=\left(e^{-\nu}, 0, \hdots, 0\right)$, the explicit form of Eq.~\eqref{orthogonal_eq} for $\sigma=r, \theta_1$ are, respectively:
\begin{eqnarray}
\left(p_d+\rho_d\right)e^{2(\lambda-\nu)}\frac{d^{2}\varsigma^{r}}{dt^2}+\left(\delta\rho_d+\delta p_d\right)\frac{d\nu}{dr}=-\frac{\partial\delta p_d}{\partial r},\label{eq_1_p}\\
\left(p_d+\rho_d\right)e^{-2\nu}r^2\frac{d^2\varsigma^{\theta_1}}{dt^2}=-\frac{\partial\delta p_d}{\partial\theta_1}.\label{eq_2_p}
\end{eqnarray}
From Eqs.~\eqref{along_eq} and \eqref{eq_13a_cuentas}, after some algebra, it is found that $\delta\rho_d$ is given by the expression:
\begin{equation}\label{eq_deltarho_cuentas}
\begin{split}
\frac{\delta\rho_d}{p_d+\rho_d}&=-\frac{{\tilde Q}}{p_d+\rho_d}\frac{d\rho_d}{dr}\frac{e^{-\lambda}}{r^{d-2}}Y_{l}^{m}\\
&-\left(\frac{e^{-\lambda}}{r^{d-2}}\frac{\partial {\tilde Q}}{\partial r}Y_{l}^{m}+\frac{{\tilde Z}}{r^{2}}l(l+d-3)Y_{l}^{m}\right).
\end{split}
\end{equation}
Considering that the fluid pressure depends on the energy density $p_d=p_d(\rho_d)$, we have:
\begin{equation}\label{eq_deltap_cuentas}
\begin{split}
\frac{\delta p_d}{p_d+\rho_d}&=-\frac{{\tilde Q}}{p_d+\rho_d}\frac{dp_d}{dr}\frac{e^{-\lambda}}{r^{d-2}}Y_l^m\\
&\hspace{-0.5cm}-\frac{dp_d}{d\rho_d}\left(\frac{e^{-\lambda}}{r^{d-2}}\frac{\partial {\tilde Q}}{\partial r}Y_l^m+\frac{{\tilde Z}}{r^2}l(l+d-3)Y_l^m\right).
\end{split}
\end{equation}
Since within the Cowling approximation the metric perturbation are neglected, the density perturbation sets to zero $\delta\rho_d=0$, however, the pressure perturbation $\delta p_d$ is not set to zero. Qualitative correct results can be derived with this approach,  see, e.g., \cite{mcdermott1983}. Using the equalities \eqref{eq_deltarho_cuentas} and \eqref{eq_deltap_cuentas} with $\delta\rho_d=0$, Eqs.~\eqref{eq_1_p} and \eqref{eq_2_p} can be written into the form:
\begin{eqnarray}
&&\hspace{-0.6cm}\frac{(p_d+\rho_d)}{r^{d-2}}e^{\lambda-2\nu}\frac{\partial^{2}{\tilde Q}}{\partial t^2}-\left(\frac{dp_d}{dr}+\frac{d\rho_d}{dr}\right)\frac{d\nu}{dr}\frac{{\tilde Q}\,e^{-\lambda}}{r^{d-2}}\nonumber\\
&&\hspace{-0.6cm}-\frac{\partial}{\partial r}\left[p_d\,\Gamma_1\left(\frac{e^{-\lambda}}{r^{d-2}}\frac{\partial {\tilde Q}}{\partial r}+\frac{{\tilde Z}}{r^{2}}l(l+d-3)\right)+\frac{{\tilde Q}e^{-\lambda}}{r^{d-2}}\frac{dp_d}{dr}\right]\nonumber\\
&&\hspace{-0.6cm}+\frac{dp_d}{dr}\left(1+\frac{dp_d}{d\rho_d}\right)\left(\frac{e^{-\lambda}}{r^{d-2}}\frac{\partial {\tilde Q}}{\partial r}+\frac{{\tilde Z}}{r^{2}}l(l+d-3)\right)=0,\label{eq_16a_cuentas}\\
&&\hspace{-0.6cm}(p_d+\rho_d)e^{-2\nu}\frac{\partial^{2}{\tilde Z}}{\partial t^2}+p_d\Gamma_1\left[\frac{e^{-\lambda}}{r^{d-2}}\frac{\partial {\tilde Q}}{\partial r}+\frac{{\tilde Z}}{r^2}l(l+d-3)\right]\nonumber\\
&&\hspace{-0.6cm}+\frac{e^{-\lambda}{\tilde Q}}{r^{d-2}}\frac{dp_d}{dr}=0, \label{eq_18a_cuentas}
\end{eqnarray}
with $\Gamma_1=\frac{p_d+\rho_d}{p_d}\frac{dp_d}{d\rho_d}$ being the adiabatic index. 

Assuming a harmonic dependence on time of the perturbative variables of the form ${\tilde Q}(t,r)= Q(r)e^{i\omega t}$ and ${\tilde Z}(t,r)=Z(r)e^{i\omega t}$, with $\omega$ being the eigenfrequency, Eqs.~\eqref{eq_16a_cuentas} and \eqref{eq_18a_cuentas} can be placed in a form more appropriate for the numerical calculation. Considering $d[{\rm Eq.~}\eqref{eq_18a_cuentas}]/dr$-[Eq.~\eqref{eq_16a_cuentas}] in Eq.~\eqref{eq_18a_cuentas}, we get:
\begin{equation}\label{eq_21_cuentas}
\frac{d{Z}}{dr}=2{Z}\frac{d\nu}{dr}-\frac{e^{\lambda}{Q}}{r^{d-2}}.
\end{equation}
On the other hand, from Eqs.\eqref{eq_18a_cuentas} and \eqref{eq_21_cuentas}, we obtain:
\begin{equation}\label{eq_19_cuentas}
\frac{d{Q}
}{dr}=\frac{d\rho_d}{dp_d}\left[\omega^2r^{d-2}e^{\lambda-2\nu}{Z}+\frac{d\nu}{dr}{Q}\right]-l(l+d-3)e^{\lambda}{Z}r^{d-4}.
\end{equation}
It is the first time that these equations are obtained, and to determine the oscillation spectrum of compact stars in a space-time in high dimensions in the Cowling approximation, it is necessary to solve Eqs.~\eqref{eq_21_cuentas} and \eqref{eq_19_cuentas}. These two differential equation are reduced to those derived in \cite{sotani2011} for a four-dimensional space-time.

To integrate Eqs.~\eqref{eq_21_cuentas} and \eqref{eq_19_cuentas} from the center ($r=0$) to the surface of the hypersphere ($r=R$), the boundary conditions have to be defined. In order to find regular solution in the center, in a similar way as realized in \cite{sotani2011,doneva}, we consider that the functions ${Q}$ and ${Z}$ take the form:
\begin{equation}\label{int_cod_ro}
{Q}=Cr^{l+d-3},\hspace{1.0cm}{Z}=-C\frac{r^{l}}{l},
\end{equation}
where $C$ represents a dimensionless constant. In turn, at the surface of the object is found:
\begin{equation}\label{ext_cod_ro}
\left[\omega^2 e^{\lambda-2\nu}{Z}r^{d-2}+\frac{d\nu}{dr}{Q}\right]_{r=R}=0.
\end{equation}

\subsection{Equation of state}

For the fluid contained in the compact object, we consider that both the $d$-dimensional energy density $\rho_{d}$ and pressure $p_{d}$ are related by a linear equation of state of form \cite{arbanil_malheiro2019}:
\begin{equation}\label{eos}
p_d = \frac{(\rho_d-d\,{\cal B}_d)}{(d-1)},
\end{equation}
where ${\cal B}_d$ is a constant. In four-dimensional space-time, Eq.~\eqref{eos} displays the MIT bag model EOS. It is well-known that this EOS represents a fluid composed of up, down, and strange quarks. In \cite{witten1984}, Witten has proposed that the strange quark matter might be the true fundamental state of strongly interacting matter. This conjecture is corroborated by Farhi and Jaffe \cite{farhi_jaffe1984} taking into account massless and non-interacting quarks.

Due to the volume, the constant ${\cal B}_d$ and the functions $\rho_d$ and $p_d$ are dimension dependent units. With the aim to have these units independent of the space-time dimension, those are used of the following form ${\cal B}_d\,G_d$, $\rho_d\,G_d$, and $p_d\,G_d$. In such a form, the units of the aforementioned variables are $[\rm MeV/fm^3]$.

In this article, following \cite{arbanil_malheiro2019}, we consider $d\,{\cal B}_d\,G_d= 240\,[\rm MeV/fm^3]$. It means that, for $d=4$, the bag constant is ${\cal B}_4= 60\,[\rm MeV/fm^3]$. It is important to say that the bag constant considered is within the hadronic mass spectroscopy interval \cite{Bordbar}, $ 60\leq{\cal B}_4\leq 90\,[\rm MeV/fm^3]$.


\section{Results}\label{results}

\subsection{Numerical method}

To analyze the extra dimensions' influence in the oscillation spectrum of compact stars, the stellar structure equations, and the nonradial oscillation equations are numerically solved from the center toward the surface of the compact object. 

Once defined the EOS, the Eqs.~\eqref{dm/dr},  \eqref{tov} and \eqref{df/f} are solved for different dimensions $d$ and central energy densities $\rho_{cd}G_d$. The numerical solutions begin by integrating Eqs.~\eqref{dm/dr} and \eqref{tov} from the center to the surface of the compact object using the fourth order Runge-Kutta method, for a value of $d$ and $\rho_{cd}G_d$.

After found the parameters $p_d\,G_d$, $\rho_d\,G_d$, $m\,G_d$ and $\lambda$, we solve Eq.~\eqref{df/f} and then the nonradial pulsation equations \eqref{eq_21_cuentas} and \eqref{eq_19_cuentas} are integrated. The whole system of equations are solved by using the shooting method as described:
\begin{itemize}
    \item In the case of Eq. \eqref{df/f}, the method starts taking into account a test value of $\nu_c$. If after the integration the condition \eqref{metric_functions} is not attained, the process is repeated until to find a $\nu_c$ that satisfy that condition. 
    
    \item The nonradial pulsation equations \eqref{eq_21_cuentas} and \eqref{eq_19_cuentas} are numerically integrated. It begins considering the correct coefficient of $\nu_c$ found, in the solution of the stellar structure equations, for a certain value of $\rho_{cd}\,G_d$ and $d$, $l=2$ and a trial value of $\omega^2$. If, at the final of the numerical solution, the condition \eqref{ext_cod_ro} is not fulfilled, $\omega^2$ is corrected until satisfied in the next integration.
\end{itemize}
The numerical method implemented in this work reproduces the results of reference \cite{doneva}, for the study of the $f$- and $p_1$-modes of a neutron star in the perfect fluid case.

\subsection{Oscillation spectrum of relativistic strange stars in a $d$-dimensional space-time}\label{homogeneous_star}

\begin{figure}[ht]
\begin{center}
\includegraphics[width=0.97\linewidth]{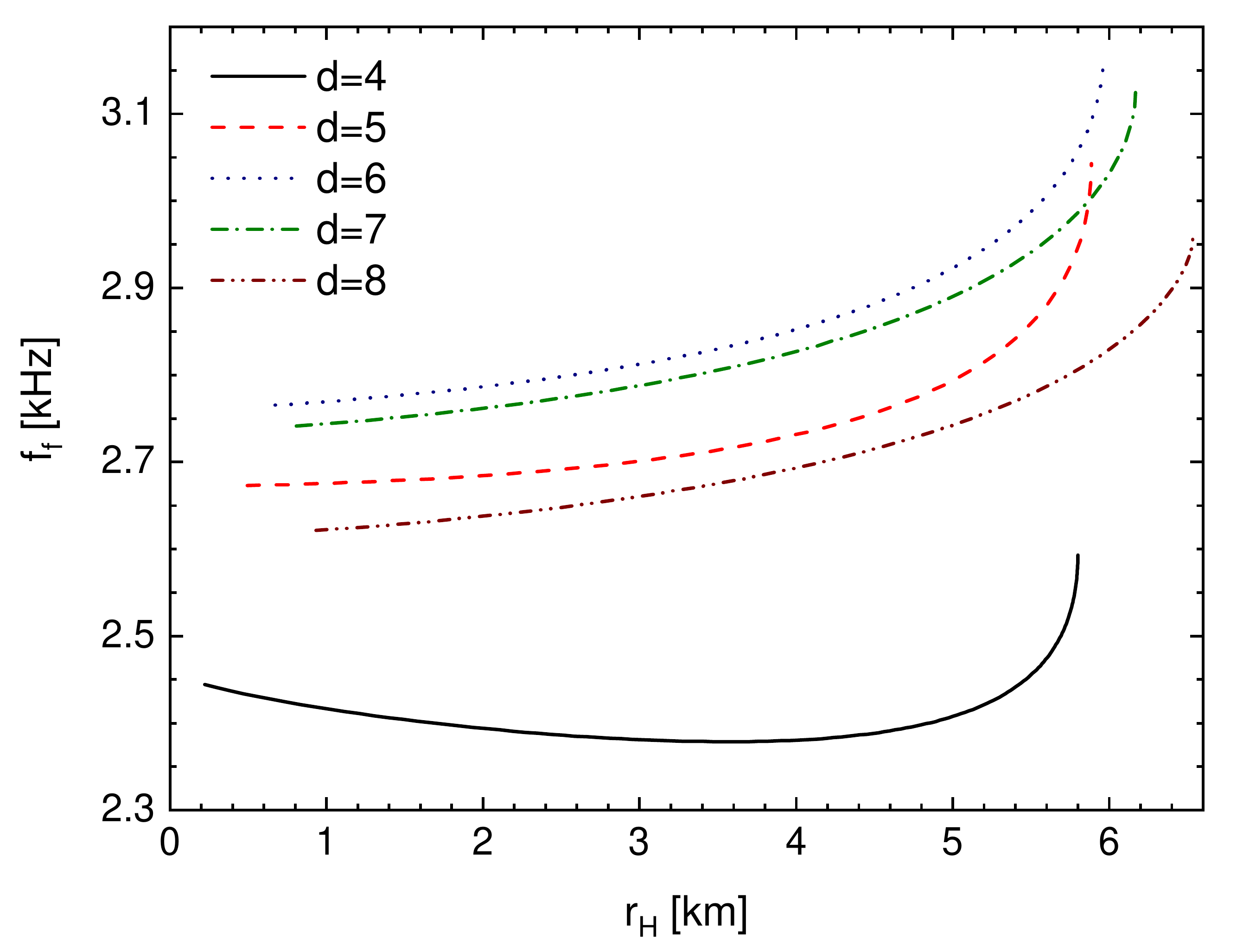}
\caption{The $f$-mode frequency versus the gravitational radius, for some space-time dimensions}
\label{ff_M_NRO}
\end{center}
\end{figure}

In Fig.~\ref{ff_M_NRO}, it is shown the behavior of the $f$-mode frequencies as a function of the gravitational radius (see, e.g., \cite{emparan_reall2008}),
\begin{equation}\label{Eq_rH}
r_H=\left(\frac{2MG_d}{d-3}\right)^{\frac{1}{d-3}}[\rm km],
\end{equation}
for some different space-time dimensions. In all cases presented, we only consider stable compact objects against radial perturbations, see \cite{arbanil_malheiro2019}. Note in the curves that the $f$-mode are almost constant \cite{sotani2003,kojima2002} and only show a fast increase for higher values of the {gravitational radius}. For $d=4$, we can see that the frequency shows a tiny small decrease with the {gravitational radius} until it reaches its minimum value. From this point, the frequency increases with $r_H$. In turn, for $d>4$, the $f$-mode is essentially constant and for large {gravitational radius} values grows monotonically with $r_H$.

\begin{figure}[ht]
\begin{center}
\includegraphics[width=0.9\linewidth]{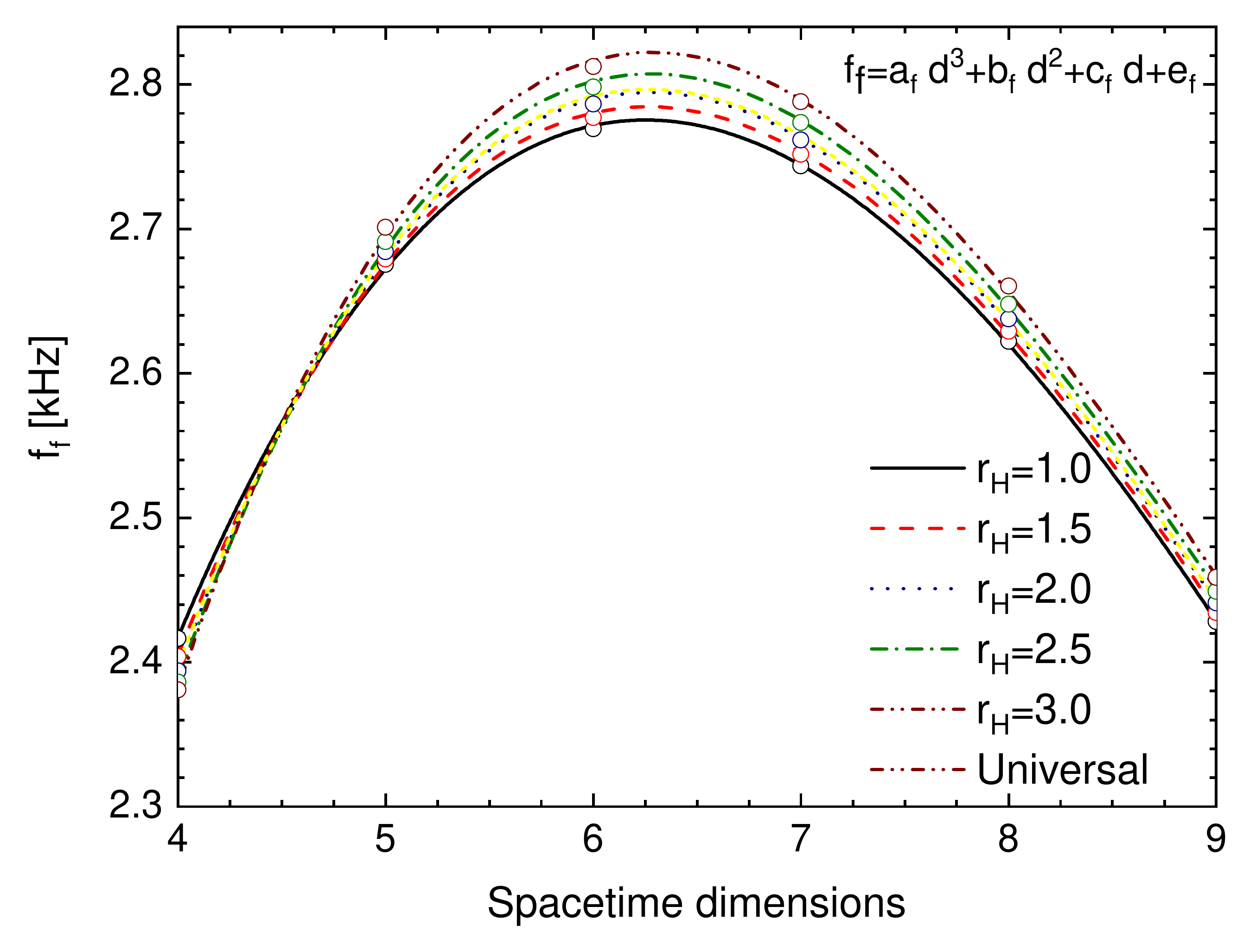}
\includegraphics[width=0.9\linewidth]{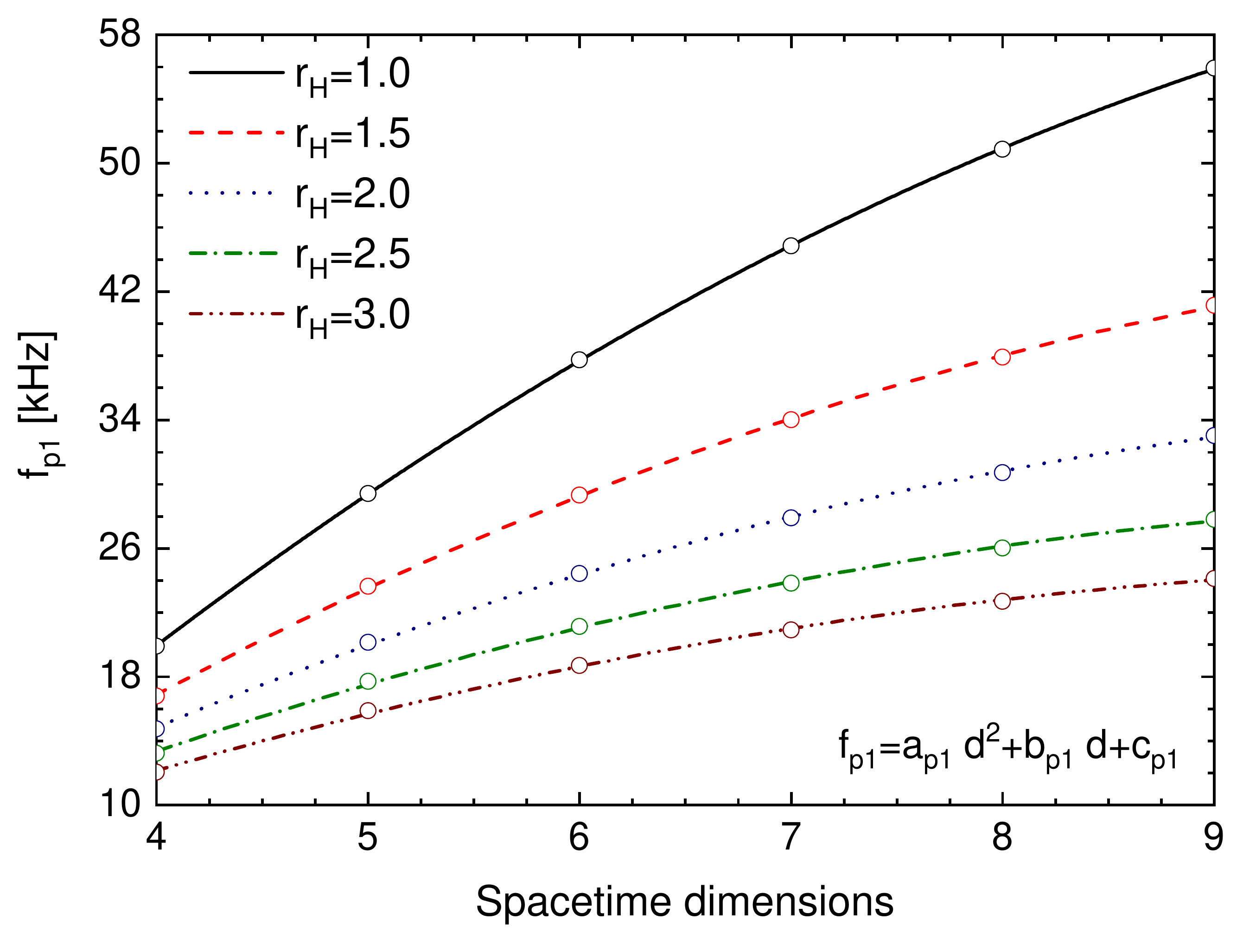}
\caption{The $f$- and $p_1$-mode frequencies as a function of the space-time dimensions are plotted on the top and bottom panel, respectively. Circles represent the frequencies values calculated with the shooting method. The curves and lines are resulting from the polynomials fitting of third and second order. In both cases are considered five different gravitational radii.}
\label{f_p1_d}
\end{center}
\end{figure}

In Fig.~\ref{ff_M_NRO} is also shown the influence of extra dimensions in the $f$-mode. For a {gravitational radius} range, we see that the $f$-mode increases in the dimensions $4\leq d\leq6$ and decreases in space-time dimensions $d\geq7$ (see also top panel of Fig.~\ref{f_p1_d}). Moreover, it is important to say that, for all space-time dimensions considered, we note that the $f$-mode frequencies are in the $2.38-3.18\,[\rm kHz]$ range. These limit values of $f$-mode frequencies are related respectively to the lowest and highest value found in the dimensions $d=4$ and $d=6$.

\begin{figure}[ht]
\begin{center}
\includegraphics[width=0.97\linewidth]{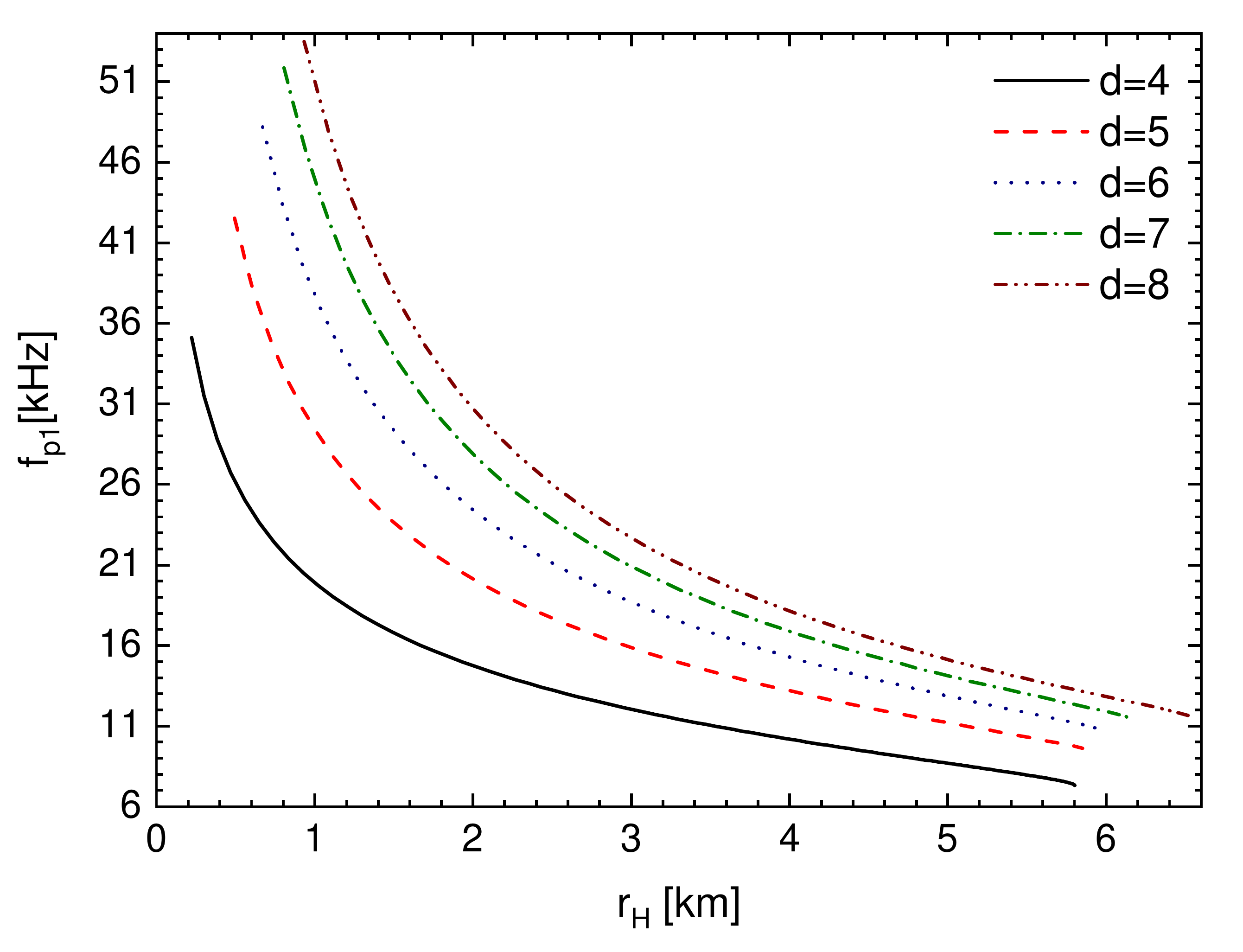}
\caption{The $p_1$-mode frequency as a function the gravitational radius, for four space-time dimensions.}
\label{fp1_M_NRO}
\end{center}
\end{figure}

The dependence of the $p_1$-mode frequencies with the {gravitational radius} for four different space-time dimensions is plotted in Fig.~\ref{fp1_M_NRO}. Such as it happens in four-dimensional space-time, in higher dimensions we found that the $p_1$-mode frequencies are larger than the $f$-modes (check Ref.~\cite{chirenti2012}). Further note that the $p_1$-mode decays monotonically with $r_H$, thus attaining the lowest frequencies of oscillation in the maximum {gravitational radius} points. On the other hand, for a {gravitational radius} interval, the $p_1$-mode frequencies are also affected by the increment of the space-time dimension. For higher dimensions larger $p_1$-modes are derived. 

The change of the $f$- and $p_1$-mode frequencies with the space-time dimensions have respectively schemed on the top and bottom of Fig.~\ref{f_p1_d}. In the figure, there are considered five different {gravitational radii}, as a function of space-time dimensions. The results obtained, marked by circles in figures (review also Table~\ref{table}), are connected by a third and second-order polynomial fitting curvature correlating $f$- and $p_1$-mode frequencies with the space-time dimensions, respectively. 

In Fig.~\ref{f_p1_d}, in the upper panel, for the range of $r_H$ chosen, the fitting curves that correlate the $f$-mode with $d$ are close to each other. This proximity in the curves is related to the interval chosen for the {gravitational radii} analyzed, where, in each dimension, the $f$-mode frequencies values are almost constant and close to each other (see Fig.\ref{ff_M_NRO}). For the mentioned {gravitational radius} range, we note that there is a universality among these curves. These can be represented by an ``universal'' curve that connects all points (review Table~\ref{tableI}). As can be noted in the figure, the $f$-mode frequency grows with the space-time dimensions until attaining its maximum value close to $d=6$, after this point, the $f$-mode decreases with the dimension. In the bottom panel, in all total {gravitational radii} considered, we obtain that the curves derived for the $p_1$-mode frequency against the number of dimensions can be reproduced quite well by a polynomial of second order in $d$. In this mode, as found in the $f$-mode, we also observe a universality among all curves $f_{p_1}=f_{p_1}(d)$  (check, again, Table~\ref{tableI}).

In Table \ref{table}, we can see the values of frequencies for $f$- and $p_1$-modes shown in Fig.~\ref{f_p1_d}, for each space-time dimension $d$ and five {gravitational radii}.

In Table \ref{tableI},  the constants of the fitting curves that correlate $f$- and $p_1$-mode frequencies with the space-time dimension in Fig.~\ref{f_p1_d}, for each $r_H$, are presented.

\begin{table*}[ht]
\begin{ruledtabular}
\caption{Values of the $f$- and $p_1$-mode frequencies for four different $r_H$. The units of the frequencies and gravitational radii are, respectively, $[\rm kHz]$ and $\left[\rm km\right]$.}
\begin{tabular}{ccccccccccc}
& \multicolumn{2}{c}{$r_H=1.0$} & \multicolumn{2}{c}{$r_H=1.5$} & \multicolumn{2}{c}{$r_H=2.0$} & \multicolumn{2}{c}{$r_H=2.5$} & \multicolumn{2}{c}{$r_H=3.0$}\\\hline
$d$ & $f$ & $p_1$ & $f$ & $p_1$ & $f$ & $p_1$& $f$ & $p_1$& $f$ & $p_1$\\\hline
$4$ & $2.4165$    & $19.910$ & $2.4039$ & $16.789$ & $2.3939$ & $14.745$ & $2.3862$ & $13.237$ & $2.3808$ & $12.042$ \\
$5$ & $2.6784$    & $29.411$ & $2.6791$ & $23.642$ & $2.6843$ & $20.141$ & $2.6913$ & $17.705$ & $2.7010$ & $15.867$ \\
$6$ & $2.7694$    & $37.747$ & $2.7771$ & $29.327$ & $2.7865$ & $24.427$ & $2.7982$ & $21.126$ & $2.8124$ & $18.703$ \\
$7$ & $2.7438$    & $44.856$ & $2.7518$ & $34.022$ & $2.7617$ & $37.844$ & $2.7736$ & $23.835$ & $2.7880$ & $20.913$ \\
$8$ & $2.6222$    & $50.863$ & $2.6291$ & $37.905$ & $2.6377$ & $30.701$ & $2.6480$ & $26.018$ & $2.6604$ & $22.681$ \\
$9$ & $2.4282$    & $55.932$ & $2.4342$ & $41.135$ & $2.4409$ & $33.022$ & $2.4490$ & $27.804$ & $2.4586$ & $24,121$
\end{tabular}\label{table}
\end{ruledtabular}
\end{table*}
        
\begin{table*}[ht]
\begin{ruledtabular}
\caption{Values of polynomial fittings parameters of the $f$- and $p_1$-mode frequencies as a function of dimensional parameter $d$, for four different $r_H$. We use third and second order polynomial fitting for $f$- and $p_1$-mode, respectively. The units of all fittings parameters are $[\rm kHz]$.}
\begin{tabular}{ccccc|ccc}
\multicolumn{5}{c}{$f-{\rm mode}$} & \multicolumn{3}{c}{$p_1-{\rm mode}$}\\\hline
$r_H$   & $a_f$ $(\times 10^{-3})$ & $b_f$ $(\times 10^{-1})$ & $c_f$ & $e_f$ & $a_{p1}$ $(\times 10^{-1})$ & $b_{p1}$ $(\times 10)$ & $c_{p1}$ $(\times 10)$ \\\hline
$1.0$         & $4.9380$ & $-1.5213$ & $1.3232$ & $-0.7575$ \ \ \ \ \ & $-5.6208$ & $1.4495$ & $-2.9042$  \\
$1.5$         & $5.5806$ & $-1.6656$ & $1.4291$ & $-1.0034$ \ \ \ \ \ & $-4.5216$ & $1.0713$ & $-1.8741$  \\
$2.0$         & $6.1148$ & $-1.7890$ & $1.5217$ & $-1.2207$ \ \ \ \ \ & $-3.7948$ & $0.8548$ & $-1.3270$  \\
$2.5$         & $6.6250$ & $-1.9080$ & $1.6118$ & $-1.4307$ \ \ \ \ \ & $-3.2789$ & $0.7133$ & $-0.9937$  \\
$3.0$         & $7.1370$ & $-2.0285$ & $1.7033$ & $-1.6420$ \ \ \ \ \ & $-2.8926$ & $0.6133$ & $-0.7746$  \\
``Universal'' & $6.0791$ & $-1.7825$ & $1.5178$ & $-1.2108$ \ \  \ \ \\
\end{tabular}\label{tableI}
\end{ruledtabular}
\end{table*}

\subsection{Oscillation spectrum of Newtonian homogeneous stars in $d$ space-time dimensions}

In this subsection, we analyze the oscillation spectrum of homogeneous stars in Newtonian gravity. We consider that the energy density $\rho_d=d\,{\cal B}_d$ is constant along whole the star. In this limit, the integration of the static equilibrium structure equations, the fluid pressure, and the mass of the object within a radius $r$ in $d$ space-time dimensions are, respectively, described by
\begin{eqnarray}
&&p_d=\frac{S_{d-2}G_d}{2(d-1)}\left(d{\cal B}_d\right)^2\left(R^2-r^2\right),\label{pressure_new}\\
&&m=\frac{S_{d-2}}{d-1}\left(d{\cal B}_d\right)\,r^{d-1}.\label{mass_new}
\end{eqnarray}
It is important to say that, within the Newtonian gravity, the non-radial oscillations in Cowling approximation, Eqs.~\eqref{eq_16a_cuentas} and \eqref{eq_18a_cuentas}, follow the relations:
\begin{eqnarray}
&&\frac{dZ}{dr}=-\frac{Q}{r^{d-2}},\\
&&\frac{dQ}{dr}=-Zr^{d-4}l\,(l+d-3).
\end{eqnarray}
From these equations, to obtain regular solutions in the center, in a similar way as considered in the relativistic case, we regard:
\begin{equation}\label{int_cod_ro_new}
{Q}=C_2r^{l+d-3},\hspace{1.0cm}{Z}=-C_2\frac{r^{l}}{l},
\end{equation}
with $C_2$ being a dimensionless constant. At the object's surface is found: 
\begin{equation}\label{ext_cod_ro_new}
\left[\omega^2{Z}r^{d-2}-\frac{dp_d}{dr}\frac{Q}{d{\cal B}_d}\right]_{r=R}=0.
\end{equation}
Considering Eqs.~\eqref{pressure_new} and \eqref{mass_new}, this last equation yields:
\begin{equation}\label{last_omega}
\omega^2=\frac{l\,M\,G_d}{R^{d-1}}.
\end{equation}
We must point out that when $l=2$ for a four-dimensional space-time $d=4$, Eq.~\eqref{last_omega} is reduced to the obtained by Emden, see, e.g., Ref.~\cite{cowling1941,gregorian2014}.

In addition, considering the relation \eqref{mass_new} for $r=R$, equation \eqref{last_omega} can be reduced to the form:
\begin{equation}\label{newton_osc}
\omega^2=l\,d\,G_d\,{\cal B}_d\frac{S_{d-2}\,}{d-1}.
\end{equation}

Through Eqs.~\eqref{last_omega} and \eqref{newton_osc}, the dependence of the eigenfrequency of oscillation $\omega$ (or the $f$-mode frequency of oscillation, since $f_f=\omega/2\pi$) with the space-time dimension can be analyzed. In that way, these two equalities are employed to investigate the dependence of the fundamental mode frequency with the squared root of the average density,
\begin{equation}
{\bar\rho}=\frac{M\,G_d}{R^{d-1}}\,[\rm km]^{-2},
\end{equation}
and different space-time dimensions. In Newtonian gravity, we can note that the $f$-mode frequency is directly proportional to ${\bar\rho}^{1/2}$ with the proportionality constant $\sqrt{l}/2\pi$; see, e.g., \cite{chirenti2012,kokkotas1999}, for $d=4$. 

In Table III, the $f$-mode of oscillations of homogeneous stars in Newtonian approximation considering $l=2$, the energy density $240\,[\rm MeV/fm^3]$, and few space-time dimensions are presented. As can be seen in Table III, the higher $f$-mode is determined in $d=6$. This is easy explained since it is well known that the volume of the unitary sphere ($S_{d-2}/(d-1)$) has a maximum for $d=6$ \cite{huber2016} and, as above mentioned, the eigenfrequency of $f$-mode oscillation (Eq.~\eqref{newton_osc}) depend on it. If we use a larger Bag constant value, consistent with hadron spectroscopy, such as $90\,[\rm MeV/fm^3]$ \cite{Bordbar} for $d=4$, that corresponds to a constant energy density of $360\,[\rm MeV/fm^3]$, the $f$-mode frequencies will increase by a factor of $\sqrt{360/240}=1.2$ because they are proportional to the squared root of the average star density that goes with ${\cal B}_4^{1/2}$ as we can see in Eq.~\eqref{newton_osc}. Thus, the $f$-mode oscillation frequencies for homogeneous stars in $d$ dimensions are still in a narrow range,  $f\sim2.8-3.3\,[\rm kHz]$.

\begin{table}[h] 
\centering
\begin{tabular}{ccc}
\hline\hline
$d$  & & $f\,[\rm kHz]$ \\\hline
$4$ & & $2.4612$ \\
$5$ & & $2.6714$ \\
$6$ & & $2.7590$ \\
$7$ & & $2.7337$ \\
$8$ & & $2.6139$ \\\hline\hline
\end{tabular}
\caption{Eigenfrequency of $f$-mode oscillation of homogeneous star in Newtonian gravity for different space-time dimensions, $l=2$ and constant energy density $240\,[\rm MeV/fm^3]$.}
\end{table}\label{tableiii}

From equation \eqref{newton_osc}, we see that for a constant $f$-mode frequency, the product $G_d d B_d S_{d-2}/(d-1)$ is equivalent to $4 G_4 B_4 S_{d-2}/(d-1)$ (since $d G_d B_d=4 G_4 B_4$). From this relation, we note that the value of the constant $f$-mode frequency for each dimension of the homogeneous star depends on the square root of the product $B_4S_{d-2}/(d-1)$. However, for any dimension, we can separate the dimensional effect from the Bag constant if we divide the constant $f$-mode frequency by the squared root of $S_{d-2}/(d-1)$. In this case, the normalized $f$-mode frequency will depend essentially only in the square root of $B_4$. Thus, if we change the value of the Bag constant in $d=4$, the new normalized $f$-mode frequency will scale according to the square root of the new value of the Bag constant in $d=4$. The dimensional effects can be separated if we calculate the ratio between the f-mode frequency in one dimension with the other in another dimension, which will depend only on the square root of the ratio between their $S_{d-2}/(d-1)$ volume factors.

\subsection{$f$-mode frequencies of relativistic strange stars and of Newtonian homogeneous stars}

\begin{figure}[ht]
\begin{center}
\includegraphics[width=0.97\linewidth]{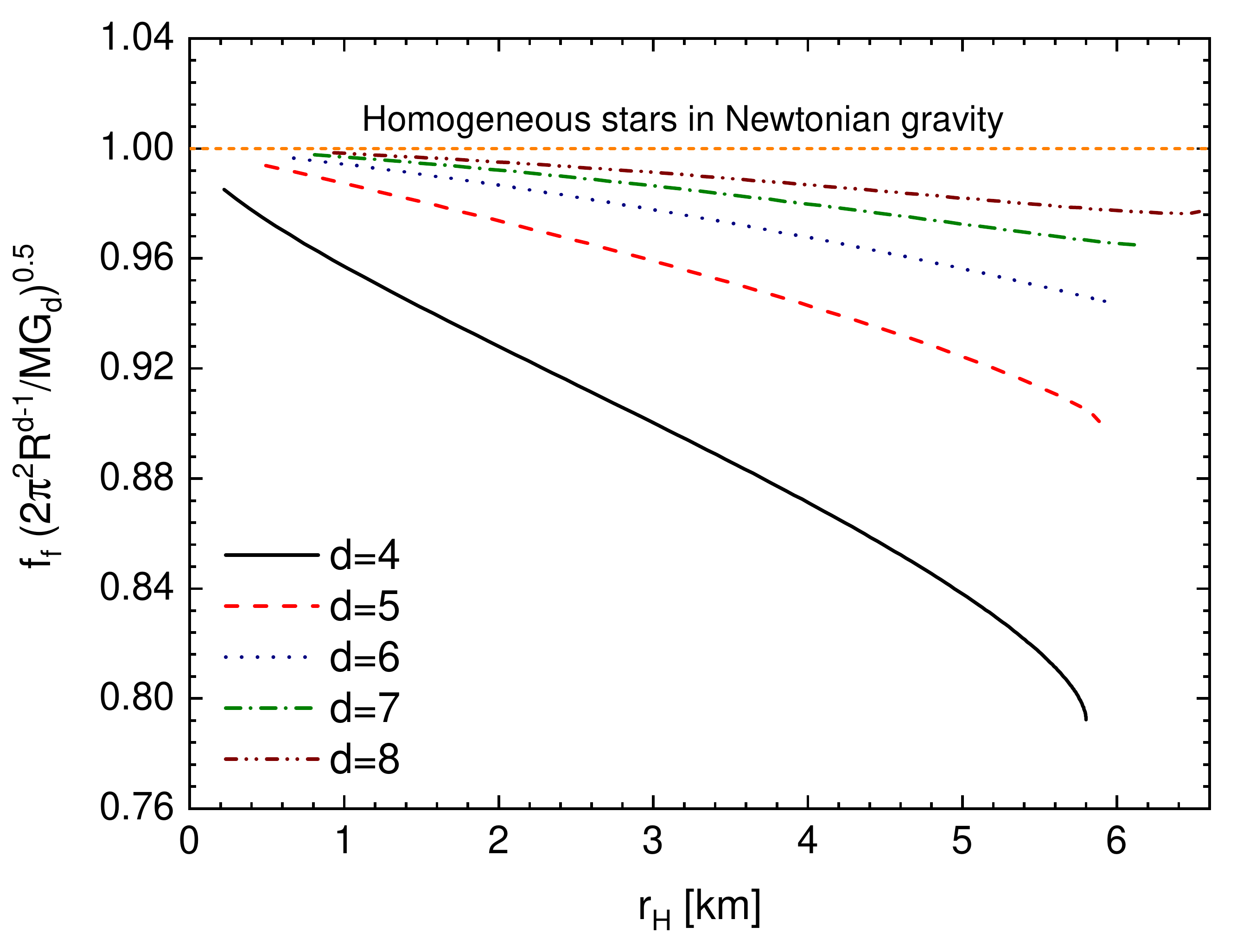}
\caption{The $f$-mode for strange quark stars in the relativistic formalism, normalized by the square root of the average density ${\bar\rho}$ divided by $2\pi^2$, i.e., $(MG_d/2\pi^2\,R^{d-1})^{0.5}$, as a function of the gravitational radius.}
\label{fNor_M}
\end{center}
\end{figure}

The main difference between Neutron stars and Quarks stars is that for these self-bound compact stars, the average density is almost constant with the variation of the total star mass for several relativistic strange quarks stars, see Fig.~\ref{fNor_M} for any dimensions, and the homogeneous density star is a good approximation. Since the star mass grows with the homogeneous energy density times the volume $S_{d-2}R^{d-1}/(d-1)$ (see Eq.~\eqref{mass_new}), the $f$-mode frequencies are almost constant for any strange quark star (independent of their mass); unlike Neutron stars, where in four-dimensional space-time, the $f$-mode frequencies scale and increase with average star density and in particular with the stellar mass. Moreover, for quarks stars with $d\,G_d\,{\cal B}_d$ fixed, we obtain $f$-mode depends only on the volume of the unitary sphere $S_{d-2}/(d-1)$. For instance, in Table III, for $d\,G_d\,{\cal B}_d=240\,[\rm MeV/fm^3]$, we find that the larger value of the $f$-mode is attained in $d=6$ ($5$ spatial dimensions).

The $f$-mode frequencies, shown in Fig.~\ref{ff_M_NRO} as a function of the gravitational radius, are almost constant and begin to increase only for a large gravitational radius. In four dimensions, we note that the $f$-mode frequency is almost constant up to $r_H\sim5.0\,[\rm km]$ (in $\sim 1.7 M_\odot$). In larger space-time dimensions, similar behavior of the $f$-mode with the gravitational radius is found, the $f$-mode remains approximately constant until $r_H\sim5.6\,[\rm km]$. In addition, in each space-time dimension, at the almost constant $f$-mode frequency interval, we note that the $f$-mode frequencies are very similar to those ones derived for homogeneous stars in Newtonian gravity, see Table III and Fig.~\ref{fNor_M}. From this, we can understand that the quark energy density is also almost constant inside the star for any number of dimensions. Additionally, for $d>6$, in the almost constant $f$-mode frequency interval, we see that the $f$-mode decreases with the space-time dimension. This is exactly the same dimensional dependence of $f$-mode frequencies of a homogeneous star in the Newtonian gravity (see, again, Table III).  This implies that the total quark star mass goes with the volume in $d-1$ dimensions and ${\bar\rho}$ is constant for any $d$ dimensions (almost constant for a self-bound quark star in $d=4$).  Due to the fact that the $f$-mode frequency is proportional to the volume of the unitary sphere, we note that it diminishes with the increase of the space-time dimension (for $d>6$). Thus, for a large dimension, the f-mode frequency tends to zero, being this the minimum value that the frequency can attain. This result seems very reasonable since the star pressure goes with $1/(d-1)$ and, due to it decreases with the space-time dimension, the star is more prone to the gravitational collapse.

\begin{table*}
    {\begin{tabular}{lcccc} 
        \hline
      SS candidate & Observed mass $M/M_{\odot}$ & Predicted radius $[\rm km]$ & $M/R$ & $f [\rm kHz]$\\
        \hline
       Vela X-$1$  & $1.77\pm0.08$ \cite{rawls/2011}  & $11.17^{+0.01}_{-0.04}$ & $0.235$  & $2.4234$\\
       $4$U $1608-52$  & $1.74\pm0.14$ \cite{guver/2010}  & $11.16^{+0.01}_{-0.12}$ & $0.231$  & $2.4164$\\
       PSR J$1903+327$ & $1.667\pm0.021$ \cite{freire/2011} & $11.11^{+0.02}_{-0.02}$ & $0.222$  & $2.4035$\\
       $4$U $1820-30$  & $1.58\pm0.06$ \cite{guver/2010a} & $11.02^{+0.07}_{-0.07}$ & $0.212$  & $2.3930$\\
       Cen X-$3$  & $1.49\pm0.08$ \cite{rawls/2011} & $10.91^{+0.02}_{-0.12}$ & $0.202$  & $2.3863$\\
       EXO $1745-248$  & $1.3\pm0.2$ \cite{ozel/2009} & $10.58^{+0.34}_{-0.35}$ & $0.182$ & $2.3794$\\
       LMC X-$4$  & $1.29\pm0.05$ \cite{rawls/2011} & $10.57^{+0.12}_{-0.11}$ & $0.181$  & $2.3791$\\
       SMC X-$1$  & $1.04\pm0.09$ \cite{rawls/2011} & $9.98^{+0.23}_{-0.25}$ & $0.154$ & $2.3804$\\
       SAX J$1808.4-3658$  & $0.9\pm0.3$ \cite{elebert/2009} & $9.73^{+0.65}_{-0.85}$ & $0.137$  & $2.3839$\\
       $4$U $1538-52$  & $0.87\pm0.07$ \cite{rawls/2011}  & $9.49^{+0.22}_{-0.12}$ & $0.136$ & $2.3852$\\
       HER X-$1$  & $0.85\pm0.15$ \cite{abu/2008}  & $9.42^{+0.48}_{-0.32}$ & $0.134$  & $2.3860$\\
       \hline
    \end{tabular}\label{tab1}}
\caption{Physical parameters of detected strange star candidates derived using the MIT EOS with ${\cal B}_4=60\,[\rm MeV/fm^3]$.}
\end{table*}

In Fig.~\ref{fNor_M}, the $f$-mode of stable relativistic quark stars -normalized by the square root of the average density ${\bar\rho}$ divided by $2\pi^2$, i.e., $(MG_d/2\pi^2\,R^{d-1})^{0.5}$- as a function of the gravitational radius. For each dimension considered, for a range of $r_H$, we see that the normalized $f$-mode follows the constant Newtonian frequencies. {At this range of $r_H$, we note that Eq.~\eqref{newton_osc} is a good approximation of the eigenfrequency squared for the relativistic case}. Furthermore, it is even more constant for higher $r_H$ and approach the homogeneous star limit when the dimension increase; only for large $r_H$ a fast decrease of the $f$-mode with the {gravitational radius} is seen. These results can be understood by analyzing the equation of state, $p_d = (\rho_d-d\,{\cal B}_d)/(d-1)$, where the fluid pressure gets much smaller than the energy density since it decreases when the space-time dimension increases, due to the factor $1/(d-1)$. Thus, for large space-time dimensions, the homogeneous star becomes a good approximation since $p_d<<\rho_d$ and to the fact that the relations $MG_d/(d-3)R^{d-3}$ \cite{arbanil_malheiro2019} and $S_{d-2}p_d\,r^{d-1}/(d-3)m$ become smaller with the grow of $d$. This indicates that, for a sufficiently large dimension, the relativistic results are similar to those ones derived from the homogeneous stars in Newtonian gravity.  From this result, as similar happen on homogeneous stars in Newtonian gravity, the $f$-mode frequency of oscillation in relativistic quark stars could have much lower values for $d>>6$.

The results of our model for $d=4$ can be compared with the parameters of strange star candidates. In Table IV, we present the mass of strange star candidates and the predicted total radius, compactness, and $f$-mode frequencies using the MIT EOS with ${\cal B}_4=60\,[\rm MeV/fm^3]$. Very different values of star masses in the range $0.85\leq M/M_\odot\leq1.77$ have $f$-mode oscillation frequencies essentially constant with values in a very small range $f\sim2.38-2.42\, [\rm kHz]$. From these results, we can understand that if $f$-mode frequencies measured by gravitational wave astronomy are almost constant, for very different pulsar masses, it is a good signature to identify strange star candidates.

\section{Conclusions and final remarks}\label{conclusion}

In this work, we investigated how the oscillation spectrum change with the space-time dimensions in the context of general relativity. Thus, we derived for the first time the nonradial perturbation equations for a $d$-dimensional space-time within the Cowling approximation. For the fluid, we consider that pressure and energy density are related by a linear EOS, the MIT bag model equation of state extended for $d$ dimensions. We also assumed that the interior solutions are connected smoothly with the vacuum Schwarzschild-Tangherlini exterior solution. We analyzed the oscillation spectrum only for stable compact objects against radial perturbations \cite{arbanil_malheiro2019} for different gravitational radii, $r_H$, and space-time dimensions, $d$. 

By analyzing the properties of the oscillation spectrum in strange quark stars, for a gravitational radius range and some space-time dimension values, it is observed that $f$-mode frequencies are almost constant and only show a fast increase for higher values of the gravitational radius, in contrast with $p_1$-mode frequencies that change significantly with  $r_H$ and $d$. From one side, for a range of gravitational radii, we found that the minimum and maximum $f$-mode frequencies were determined on the dimensions $4$ and $6$, respectively. Moreover, for a range of gravitational radii, the $f$- mode frequency dependence with the number of dimensions $d$ follows an universal curve as a polynomial of third order in $d$. On the other hand, at the same gravitational radii interval, we found that the $p_1$-mode frequencies grow significantly with $d$, and the frequency dependence with the number of dimensions can be reproduced quite well with a polynomial of second order in $d$. As happens in $d=4$, for $d>4$, we found that the $f$-mode frequencies are lower than those ones obtained for $p_1$-mode.

Within the Cowling approximation,  the nonradial oscillation equations in the Newtonian gravity were also analyzed. For this framework, we investigated the oscillation spectrum of a compact object made of homogeneous energy density, for such a case, it is considered $\rho_d=d\,{\cal B}_d$=cte. along whole star. Such as it is found in the four-dimensional case, for a $d$-dimensional space-time, the eigenfrequency of oscillation of a compact object also depends on its total mass $M\,G_d$ and radius $R$.  Furthermore, since quark star mass goes with the volume in $d$ dimensions, such as it happens in $d=4$, the eigenfrequency squared of the $f$-mode in the Newtonian gravity is constant and depend only on the volume of the unitary sphere ($S_{d-2}/(d-1)$), which has a maximum for $d=6$. 

We prove relativistic $f$-mode frequencies for $d\geq4$, that we have calculated here for the first time, are almost constant for any strange quark star independent of their masses if $M\leq 1.8 M_\odot$. From this, for the gravitational radius range where the frequencies are essentially constant, we can also note that quark energy density is also almost constant inside the star for any number of dimensions. This implies that for relativistic strange stars, for any $d$ dimensions, the total quark star mass still goes approximately with the volume in $d-1$ dimensions and the average star density ${\bar\rho}$ that is almost constant as a function of the total mass.

It is important to stress that for neutron stars in $d=4$ dimensions, the star mass increase when the stellar radius decrease and, as a consequence the average density increases with the central density and is not constant, which implies that $f$-mode frequencies increase with the neutron star mass \cite{lugones2014}. Thereby, the possibility of measure in gravitational wave detectors the $f$-mode oscillation frequency emitted by compact stars \cite{alford2019,glampedakis2019,Ho2020}, for different pulsar masses, and obtain almost constant frequency values, for $d=4$, in the interval $f\sim 2-3\,[\rm kHz]$ with $ M\leq 1.7 M_\odot$, it would be a good indication of the existence of strange quark stars which still lack an astronomical confirmation. Finally, if the $f$-mode frequencies are still constant and larger than those values found in $d=4$ for a range of larger total masses, it would be a sign that quarks can propagate in extra space-time dimensions and strange quarks stars in  $d$ dimension could exist. 

It is worth noting that strange quark stars with crust are also investigated in the literature \cite{lattimer}. The crust on the top of such stars could be supported by strong electric fields at the surface. On that work, it is concluded that the existence of a crust results in large radii for small stellar masses  $\sim 0.01 M_\odot$, but it does not considerably affect the radii of quark stars with larger masses. Therefore, the existence of a crust does not seem to have an important effect on the quadrupole properties of the star \cite{lattimer,yip} for quark star masses larger than one solar mass. From these findings, our conclusion of $f$-mode oscillation frequencies being almost constant for very different quark star masses should be still valid for strange quark stars with the crust. This could be a good signature to confirm the existence of strange quark stars, that still lack an astronomical confirmation, and might be found in gravitational-wave astronomy \cite{Ho2020}.

\begin{acknowledgments}
\noindent We would like to thank Funda\c{c}\~ao de Amparo \`a Pesquisa de S\~ao Paulo-FAPESP, under the thematic project Grant No. $2013/26258-4$. M.M. is also grateful to Conselho Nacional de Desenvolvimento Cient\'ifico e Tecnol\'ogico-CNPq and Coordena\c{c}\~ ao de Aperfei\c{c}oamento de Pessoal de N\'ivel Superior-CAPES for financial support.
\end{acknowledgments}

\appendix*

\section{Divergence of the velocity}

Within the Cowling approximation, the divergence of the velocity $\nabla_{\nu}\left(\delta U^{\nu}\right)$ is derived through the equality:
\begin{equation}
\nabla_{\nu}\left(\delta U^{\nu}\right)=\frac{1}{\sqrt{-||g||}}\partial_{\nu}\left(\sqrt{-||g||}\,\delta U^{\nu}\right),
\end{equation}
with $\sqrt{-||g||}$ being of the form:
\begin{equation}
\sqrt{-||g||}=e^{\lambda+\nu}r^{d-2}\prod_{i=1}^{d-3}(\sin\theta_i)^{d-2-i}.
\end{equation}
The divergence of the velocity is represented by the equality
\begin{eqnarray}\label{eq_9_cuentas}
&&\nabla_{\nu}\left(\delta U^{\nu}\right)=\left(\frac{d\lambda}{dr}+\frac{d\nu}{dr}+\frac{d-2}{r}\right)\delta U^r+\partial_r\left(\delta U^r\right)\nonumber\\
&&+\sum_{i=1}^{d-2}\partial_{\theta_i}\left(\delta U^{\theta_i}\right)+\sum_{i=1}^{d-3}(d-2-i)\frac{\delta U^{\theta_i}}{\tan\theta_{i}}.
\end{eqnarray}

In a similar way as defined in four dimensions, for a higher-dimensional space-time, Eq.~\eqref{int_match}, the spherical harmonic follows the identity \cite{frye2012}:
\begin{equation}\label{eq_12_2_cuentas}
\Delta_{S^{d-2}}Y_{l}^{m}=-l(l+d-3)Y_{l}^{m},
\end{equation}
where, following Ref.~\cite{umemura1965}, $\Delta_{S^{d-2}}$ is given by the expression:
\begin{equation}
\Delta_{S^{d-2}}=\sum_{i=1}^{d-2}\prod_{j=1}^{i-1}\frac{\partial^{2}_{\theta_i}}{\sin^2\theta_{j}}+\sum_{i=1}^{d-3}\prod_{j=1}^{i-1}(d-2-i)\frac{\cot\theta_{i}\partial_{\theta_i}}{\sin^2\theta_{j}}.
\end{equation}
Taking into account the fluid velocity component, Eq.~\eqref{uo}, after some algebra Eq.~\eqref{eq_12_2_cuentas} is reduced to the form:
\begin{equation}\label{eq_13_cuentas}
\sum_{i=1}^{d-2}\partial_{\theta_i}(\delta U^{\theta_i})+\sum_{i=1}^{d-3}(d-2-i)\frac{\delta U^{\theta_i}}{\tan\theta_{i}}=\frac{l(l+d-3)}{r^2 e^{\nu}}\frac{\partial{\tilde Z}}{\partial t}Y_{l}^{m}.
\end{equation}
Replacing Eqs.~\eqref{ur} and \eqref{eq_13_cuentas} in Eq.~\eqref{eq_9_cuentas}, it becomes:
\begin{equation}\label{eq_13a_cuentas}
\nabla_{\nu}\left(\delta U^{\nu}\right)=\frac{e^{-(\lambda+\nu)}}{r^{d-2}}\frac{\partial^2{\tilde Q}}{\partial t\partial r}Y_{l}^{m}+\frac{l(l+d-3)}{r^2 e^{\nu}}\frac{\partial {\tilde Z}}{\partial t}Y_{l}^{m}.
\end{equation}

\end{document}